\documentclass{article}
\usepackage{graphicx} 
\usepackage{dcolumn}
\usepackage{bm}
\usepackage{physics}
\usepackage[utf8]{inputenc}
\usepackage[T1]{fontenc}
\usepackage{mathptmx}
\usepackage{etoolbox}

\title{Exploring Nonlinear Drift Waves: Limiting Cases and Dynamics}
\author{H. Saleem$^{1,2,3,*}$\\
$^1$ Theoretical Research Institute, \\ Pakistan Academy of Sciences (TRIPAS), 3-Constitution Avenue, \\ G-5/2, Islamabad (44000), Pakistan.\\
$^2$ Department of Physics and Astronomy, School of Natural Sciences (SNS), \\ National University of Sciences and Technology (NUST), \\ H-12, Islamabad (44000), Pakistan.\\
$^3$ Department of Scpace Science, Institute of Space Technology (IST), \\ 1-Islamabad Expressway, Islamabad (44000), Pakistan.\\
$^*$ saleemhpk@hotmail.com
}
\date{January 2026}

\begin{document}

\maketitle

\begin{abstract}
A general equation for drift waves is derived incorporating both nonlinear electron density perturbation and ion vorticity effects. It is emphasized that the well-known Hasegawa–Mima (HM) equation for drift waves [A. Hasegawa and K. Mima, Phys. Fluids 21, 87 (1978)] includes only the ion vorticity term and neglects nonlinear electron density contribution that naturally arises from the electrons Boltzmann response. If ion vorticity term is ignored, then the general nonlinear equation reduces to an equation which can give two-dimensional soliton solution under an appropriate coordinate transformation. Furthermore, under the assumption that the normalized electrostatic potential depends only on one spatial coordinate along the predominant propagation direction, i.e. $\Phi = \Phi(y)$, the equation reduces to one-dimensional KdV equation [H. Saleem, Phys. Plasmas 31, 112102 (2024)]. Conversely, if the nonlinear electron density term is artificially suppressed and a two-dimensional potential $\Phi = \Phi(x, y)$ is considered, the equation reduces to Hasegawa–Mima equation supporting dipolar vortex solution. Because the HM equation ignores nonlinear electron density term, it cannot support one- or two-dimensional soliton solutions. Finally, the limiting forms of the general nonlinear equation are also briefly discussed using the reductive perturbation method (RPM).
\end{abstract}
Key Words: Nonlinear Drift Waves, Hasegawa-Mima Equation, Korteweg de-Vries Equation, Dip Solitons, Reductive Perturbation Method.

\section{Introduction}
The drift wave is an important low frequency electrostatic normal mode of magnetized inhomogeneous plasma \cite{rudakov1960oscillations, kadomtsev1962drift}. It has invoked the interest of plasma physics community because of its relevance to cross-field particle and energy transport as well as the turbulence \cite{horton1999drift, weiland1999collective, krasheninnikov2024anomalous}. Given that most magnetized plasmas—whether in space \cite{sato1989space}, astrophysical settings \cite{priest2014magnetohydrodynamics}, or laboratory environments \cite{weiland1999collective, knyazev2024resistive}—exhibit inherent inhomogeneities, the drift wave has attracted considerable research interest since its initial discovery \cite{rudakov1960oscillations, kadomtsev1962drift, d1963low}. In collisionless cold-ion plasmas, it remains a stable mode under the framework of fluid theory; however, the inclusion of electron-ion collisions renders it unstable. In such collisional regimes, this instability is referred to as the drift dissipative instability \cite{chen2012introduction}.

This wave is often called the universal instability because kinetic theory predicts that it can draw energy from electrons, becoming unstable via wave–particle interactions. A field-aligned electron current can further enhance this instability \cite{Bellan2006, Ichimaru2018}, even in the absence of collisions. By contrast, two-fluid theory indicates that a field-aligned ion shear flow can also drive unstable drift waves in cold, collisionless plasmas \cite{Saleem2007a}. Drift waves have likewise been investigated in hot plasmas using multi-fluid models to explore their potential role in self-heating of the solar corona \cite{Saleem2012}. 

Nonlinear drift waves can give rise to coherent structures such as vortices, as predicted by the Hasegawa–Mima (HM) equation \cite{hasegawa1978pseudo}, as well as shocks \cite{tasso1967shock} and solitons \cite{nozaki1974propagation, Saleem2024}. Both linear and nonlinear drift waves have recently been investigated in multi-component space plasmas \cite{shan2024coupled}. Moreover, several authors have sought to explain solar coronal heating via drift wave theory formulated in a kinetic framework applicable to the hot solar corona \cite{Vranjes2009a, Vranjes2009, Brchnelova2024}.

It is important to note that the well-known Hasegawa–Mima equation \cite{hasegawa1978pseudo} ignores the nonlinear contribution arising from Boltzmann density response of inertia-less electrons. To address this issue, we derive a more general nonlinear equation for drift waves, allowing direct comparison of the relative strengths of the scalar nonlinearity (arising from Boltzmann electrons) and vector nonlinearity (associated with ion vorticity). Our results show that the scalar nonlinear density term generally dominates over the vector nonlinearity. Furthermore, if the weaker nonlinear term is neglected under appropriate conditions, we recover the result derived in Ref. \cite{Saleem2024}.

It has been demonstrated \cite{Saleem2024} that the Korteweg–de Vries (KdV) equation for drift waves should be derived by normalizing the ion velocity with the electron diamagnetic drift speed, which matches the true drift wave phase speed. In that derivation, the parallel component of ion velocity is omitted, as in the HM equation \cite{hasegawa1978pseudo}. However, retaining parallel ion motion introduces coupling between drift and ion acoustic waves \cite{kadomtsev1965plasma}, and the resulting dispersion relation yields a phase speed close to the ion acoustic speed. Earlier  derivations of KdV equation for these waves included the parallel ion motion and the ion velocity was normalized with the ion acoustic speed \cite{nozaki1974propagation, gell1977drift}. Under this latter normalization, the KdV–Burgers equation for collisional plasmas was also derived \cite{goswami1977finite}. 
In an early investigation \cite{Oraevskii1969}, it was proposed that the drift waves produce solitary structures if the electron temperature gradient is non-zero. Later on, Lakhin et al. \cite{lakhin1987revision, lakhin1988drift} showed that the drift solitons could be produced in the presence of only density gradient and electron temperature gradient was not necessary. About a decade ago \cite{krasheninnikov2016origin, zhang2016blobs}, the plasma density blobs were investigated using nonlinear drift wave theory. In Refs. \cite{krasheninnikov2016origin, zhang2016blobs}, the nonlinear electron density contribution was taken into account and one-dimensional energy integral equation was derived to discuss drift solitons. On the other hand, in the present work, we explore the two dimensional drift wave propagation including the effects of both ion vorticity and the nonlinear electron density ignoring the propagation along field lines and assuming the density gradient scale length to be constant following Hasegawa-Mima \cite{hasegawa1978pseudo}. 

A key insight from Ref. \cite{Saleem2024} is that drift waves propagate predominantly perpendicular to both the density gradient and the background magnetic field, with a phase velocity equal to the electron diamagnetic drift velocity. Hence, to describe nonlinear drift waves accurately, the parallel ion velocity must be neglected. Inertia-less electrons, moving along magnetic field lines, produce a scalar nonlinear density response that is not canceled by ion parallel current in the drift wave regime. This imbalance leads to the formation of density cavities, in contrast to the density humps predicted by the KdV equation for ion acoustic waves \cite{washimi1966propagation}. Finally, we revisit the nonlinear dynamics of drift waves to obtain a more general equation.


\section{Nonlinear equation for drift waves}
First, we just mention the simplest drift wave geometry in Cartesian coordinates to lay down the background for highlighting the missing scalar nonlinearity in Hasegawa-Mima (HM) equation \cite{hasegawa1978pseudo}. If we choose the  ambient magnetic field to be along z-axis ${\bf B}_0=B_0 \hat{z}$, the density gradient along negative x-axis i.e. $\nabla n_0 = - \hat{x} \mid \frac{d n_0}{d x} \mid$ and density gradient scale length $L_n = \kappa_n ^{-1}=$ constant with $\kappa_n= \mid \frac{1}{n_0} \frac{d n_0}{d x} \mid$, then the linear drift wave propagates predominantly along y-direction with $k_z<<k_y$. Wave propagation is considered to be in yz-plane with wave vector ${\bf k} = (0,k_y,k_z)$. The small but non-zero component of wave vector along field lines allows the inertia-less electrons to follow the Boltzmann density distribution in wave's oscillating electrostatic potential. The conditions for this wave to propagate in a collision-less plasma can be summarized as follows, 
\begin{equation}\label{wave prop cond}
	\omega \ll \Omega_i;\quad v_{Ti} k_z \ll \omega \ll v_{Te} k_z;\quad k_z\ll k_{y};\quad \kappa_n \ll k_{y}
\end{equation} 
where $v_{Ti}^2=\frac{T_i}{m_i}$, $v_{Te}^2=\frac{T_e}{m_e}$, and $\Omega_i=\frac{eB_0}{m_i c}$. For Fourier analysis, we need the local approximation to be valid i.e. $\kappa_n \ll k_{y}$.
Since its a low frequency wave, therefore the inertial-less electrons ($m_e \rightarrow 0$) are assumed to follow the Boltzmann distribution,
\begin{equation}
	\frac{n_e}{n_{e0}} \simeq \exp({\frac{e \varphi}{T_e}}) \simeq 1 + \Phi + \frac{1}{2} \Phi^2 + ...
\end{equation}
where $\Phi= \frac{e \varphi}{T_e}$. The perpendicular component of cold ion velocity in drift approximation $\omega << \Omega_i$ is expressed in the following form,
\begin{equation}
	{\bf v}_{i \perp} = - \frac{c}{B_0} \nabla_{\perp} \varphi \times \hat{z} -\frac{c}{B_0 \Omega_i}(\partial_t + {\bf v}_i \cdot \nabla) \nabla_{\perp} \varphi ={\bf v}_E+{\bf v}_{P}
\end{equation}
where ${\bf v}_E$ is the electric drift, ${\bf v}_{P}$ is ion polarization drift and $\nabla_{\perp}=(\partial_x, \partial_y)$.
It gives,
\begin{equation}
	\nabla_{\perp} \cdot {\bf v}_{i \perp} \simeq -\frac{c}{B_0 \Omega_i}(\partial_t + {\bf v}_E \cdot \nabla) \nabla_{\perp}^2 \varphi
\end{equation}
where $\nabla_{\perp} \cdot {\bf v}_E=0$. 
Recently \cite{Saleem2024}, it has been pointed out that the scalar nonlinearity which is represented by the term $\frac{1}{2} \varphi^2$ on right hand side (rhs) of relation (2) plays an important role in creating dip solitons. Using reductive perturbation method (RPM), the KdV equation for drift waves has been obtained in the following form,
\begin{equation}
	\partial_{\tau} \Phi - \Phi \partial_{\xi} \Phi + \partial_{\xi}^3 \Phi =0
\end{equation}
The above KdV equation for drift waves has the form similar to the KdV equation for ion acoustic waves (IAWs) with an important difference that the nonlinear term in drift wave equation is negative and it can produce density cavities which have been observed in upper ionosphere \cite{dovner1994freja, cairns1995electrostatic}. 

The KdV equation for drift waves has been obtained assuming that the physical quantities associated with this wave are functions of only one spatial coordinate i.e. $\varphi= \varphi (y)$ and in this situation, Eq. (4) reduces to,
\begin{equation}
	\nabla_{\perp} \cdot {\bf v}_{i \perp} \simeq -\frac{c}{B_0 \Omega_i}(\partial_t ) \nabla_{\perp}^2 \varphi
\end{equation}
The above relation gives the dispersive term in KdV equation (5) and nonlinear term is the contribution of nonlinear electron density perturbation expressed in terms of electrostatic potential in Eq. (2).

Now we discuss the well-known Hasegawa-Mima (HM) equation for drift waves which admits the solutions in the form of dipolar vortex. Let us consider the ion continuity equation,
\begin{equation}
	\partial_t n_i+ \nabla_{\perp} \cdot (n_i {\bf v}_i)=0
\end{equation}
which can be expressed more explicitly as,
\begin{equation}
	\begin{aligned}
		&(\partial_t + {\bf v}_E \cdot \nabla_{\perp}) \frac{\tilde{n_i}}{n_{i0}} + \frac{\nabla n_{i0}}{n_{i0}} \cdot {\bf v}_E + \nabla_{\perp} \cdot {\bf v}_P \\&
		+ [\frac{\tilde{n_i}}{n_{i0}}\nabla_{\perp} \cdot {\bf v}_P+\frac{\nabla n_{i0}}{n_{i0}} \cdot {\bf v}_P+ \frac{\nabla \tilde{n_i}}{n_{i0}} \cdot {\bf v}_P] =0
	\end{aligned}
\end{equation}
where $n_i = n_{i0}+ \tilde{n_i}$ and $\tilde{n_i}$ is the perturbed part of ion density. If physical variables are assumed to be the functions of $(x,y)$ coordinates i.e. $\varphi=\varphi(x,y)$, then vorticity associated with drift wave dynamics becomes nonzero because in this case $\nabla \times {\bf v}_E= \frac{c}{B_0}\nabla_{\perp}^2 \varphi (x,y) \hat{z} \neq 0$. 
The electric drift is larger than the polarization drift due to $\omega<<\Omega_i$. The terms in square bracket of Eq. (8) are smaller compared to the term $\nabla_{\perp} \cdot {\bf v}_P$ and hence it reduces to,
\begin{equation}
	(\partial_t + {\bf v}_E \cdot \nabla_{\perp}) \frac{\tilde{n_i}}{n_{i0}} + \frac{\nabla n_{i0}}{n_{i0}} \cdot {\bf v}_E + \nabla_{\perp} \cdot {\bf v}_P \simeq 0
\end{equation} 
Using Eqs. (3) and (9) along with Boltzmann distribution for electrons, the Hasegawa-Mima equation can be written in the following form,
\begin{equation}
	\partial_t(\Phi - \rho_s^2 \nabla_{\perp}^2 \Phi) + D_e (\nabla_{\perp} \Phi \times \hat{z} \cdot \nabla) [\rho_s^2 \nabla_{\perp}^2 \Phi + \ln \frac{1}{n_{i0}}]=0
\end{equation}
where $\rho_s=\frac{c_s}{\Omega_i}$ and $D_e=\frac{cT_e}{e B_0}$.
Equation (10) is the same as Eq. (2.9.9) of Ref. \cite{sato1989space}. This equation has been obtained by replacing $\frac{\hat{n_i}}{n_{i0}}$ with $\Phi$ in the first term on left hand side (lhs) of Eq. (9). Note that, 
\begin{equation}
	(\partial_t + {\bf v}_E \cdot \nabla_{\perp}) \frac{\tilde{n_i}}{n_{i0}} = (\partial_t + {\bf v}_E \cdot \nabla_{\perp}) \Phi
\end{equation}
was considered and therefore the scalar nonlinear term of relation (2) i.e.  $(\frac{1}{2} \varphi^2)$ did not appear in the HM equation. 

Now we derive a general nonlinear equation for drift waves which will reduce to KdV equation (5) in one dimensional case for $\varphi= \varphi (y)$. 
In two dimensional case $\Phi=\Phi(x,y)$, it will reduce to HM like equation if scalar nonlinear term is dropped artificially. Furthermore, it will yield a two dimensional soliton solution in a limiting case as well.\\
To normalize Eq. (9), we define $n_i^{\prime}=\frac{n_i}{n_0}$, ${\bf v}_E^{\prime}= \frac{{\bf v}_E^{\prime}}{v_{De}}$, $y^{\prime}=\frac{y}{\rho_s}$, $x^{\prime}=\frac{x}{\rho_s}$, $\nabla^{\prime}= \rho_s \nabla$ and $t^{\prime}=\frac{c_s}{L_n}t$ where $\frac{c_s}{L_n}=\frac{v_{De}}{\rho_s}$, $v_{De}= D_e \kappa_n$ is the electrons diamagnetic drift velocity. 
Equation (9) in normalized form becomes,
\begin{equation}
	\partial_t (\frac{\tilde{n_i}}{n_{i0}}- \nabla_{\perp}^2 \Phi) + (\nabla_{\perp} \Phi \times \hat{z}. \hat{x}) + \frac{L_n}{\rho_s} (\nabla_{\perp} \Phi \times \hat{z} \cdot \nabla_{\perp}) \nabla_{\perp}^2 \Phi =0
\end{equation}
where the superscript prime (${\prime}$) has been dropped for simplicity. It may be mentioned here that $\frac{\tilde{n}_i}{n_{e0}}=(\Phi + \frac{1}{2} \Phi^2)$ and  $\frac{\tilde{n}_i}{n_{e0}} \neq \Phi$.
The above equation can be expressed in terms of a single variable $\Phi$ as follows,
\begin{equation}
	\partial_t (\Phi + \frac{1}{2} \Phi^2- \nabla_{\perp}^2 \Phi) + (\nabla_{\perp} \Phi \times \hat{z}. \hat{x}) + \frac{L_n}{\rho_s} (\nabla_{\perp} \Phi \times \hat{z} \cdot \nabla_{\perp}) \nabla_{\perp}^2 \Phi =0
\end{equation} 
The Eq. (13) is a new nonlinear equation for two-dimensional electrostatic drift waves propagating in a plane perpendicular to ${\bf B}_0$.This equation will also be expressed in stretched coordinates in section IV and the magnitudes of different terms will be compared. Before doing that, we would like to present a few comments on the HM equation in the next section.

\section{Comments on Hasegawa-Mima Equation}
In this section, we elaborate the difference in the mathematical approaches adopted to derive HM Eq. (10) and the more general Eq. (13) for the nonlinear drift waves. In deriving Eq. (10), the ion continuity equation is expressed in the following form,
\begin{equation}
\frac{(\partial_t + {\bf v}_i \cdot \nabla) n_i}{n_i} + \nabla \cdot {\bf v}_i=0
\end{equation}
Let us write the time derivative term and other terms separately,
\begin{equation}
\frac{\partial_t n_i}{n_i} + \frac{{\bf v}_E \cdot \nabla (n_0(x)+\hat{n_i})}{(n_0+\hat{n_i})}	+ \nabla \cdot {\bf v}_i=0
\end{equation}
where ${\bf v}_i\simeq {\bf v}_E$ has been approximated and $\hat{n}_i=\hat{n}_i(x,y,t)$. Since $n_i\simeq n_e$, therefore in the derivation of HM equation, the first term is expressed as $\frac{\partial_t n_i}{n_i}= \partial_t \ln e^{\Phi}=\partial_t \Phi$, then the  above equation takes the following form,
\begin{equation}
\partial_t \Phi + {\bf v}_E \cdot \nabla \ln n_0 + \nabla \cdot {\bf v}_i=0
\end{equation}
This equation can be expressed in the form of Eq. (10) which is the famous HM equation.

In our point of view, the term $\frac{\partial_t n_i}{n_i}$ in Eq. (15)  requires a careful analysis. Note that
\begin{equation}
\frac{\partial_t n_i}{n_i}=\frac{\partial_t (n_0+\hat{n}_i)}{(n_0+\hat{n}_i)}= \frac{\partial_t \hat{n}_i }{(n_0+\hat{n}_i)} \neq \partial_t \ln \frac{n_i}{n_0}
\end{equation}
since $\partial_t n_0=0$. The problem arises because the term $\partial_t n_i$ is normalized with a variable total density while it must be normalized with the background density $n_0 (x)$ which is independent of time. This is a usual procedure to study the nonlinear dynamics of low frequency electrostatic waves which has also been followed by Washimi and Tanuity \cite{washimi1966propagation} in deriving KdV equation for IAWs in unmagnetized plasma. They used the following approximation in the derivation of their nonlinear equation for IAWs\cite{washimi1966propagation},
\begin{equation}
\partial_t(\frac{n_i}{n_0}) =\partial_t \tilde{n}_i \simeq (\Phi + \frac{1}{2} \Phi^2)
\end{equation}
It may be mentioned here that if the ion continuity equation is used following HM approach, then KdV equation for drift waves \cite{Saleem2024} cannot be obtained. It is also important to note that the form of well-known KdV equation for nonlinear ion acoustic waves derived by Washimi and Tanuity \cite{washimi1966propagation} will change. The nonlinear electron density term $\frac{1}{2} \varphi^2$ is canceled by the effects of ion parallel motion and the net result is that the nonlinear term in KdV equation for IAWs become positive giving rise to hump soliton contrary to dip soliton in case of drift waves.

 Let us consider the ordering of dimensionless parameters $\frac{\rho_s}{L_n}$ and $\epsilon$. The HM equation has been derived assuming $\frac{\rho_s}{L_n} \simeq \epsilon \simeq \Phi$ and $\rho_s^2 k_{\perp}^2 \simeq 1$. In this case, the nonlinear ion vorticity term seems to be comparable to the linear term $(\nabla_{\perp} \Phi \times \hat{z}. \hat{x})$ in Eq. (13) and it is concluded that the drift waves behave nonlinearly even at a very small amplitude. However, it should be noted that the drift wave is a very small frequency wave with the condition $\omega << \Omega_i$ and ion vorticity term is the contribution of small ion polarization drift. The dispersion term $\rho_s^2 k_{\perp}^2$ is usually small and fulfills the condition $\rho_s^2 k_{\perp}^2<1$. Furthermore, the condition $\mid \rho_s^2 k_{\perp}^2 \Phi  \mid  > \mid \rho_s^2 \nabla^2 \Phi  \mid $ generally holds and hence the nonlinear vorticity term is smaller than the linear term. Therefore,
the scalar nonlinearity represented by the term $\partial_t ( \frac{1}{2} \Phi^2)$ cannot be ignored and formation of one dimensional (1D) and two dimensional (2D) solitons by drift waves cannot be ruled out. The general equation (13) for two dimensional nonlinear drift waves contain both the nonlinear terms. We will compare these two nonlinear terms of Eq. (13) using reductive perturbation method in section-IV.  
\section{Nonlinear Equation in Stretched Coordinates}
Here we try to obtain nonlinear drift wave equation using reductive perturbation method (RPM) in stretched coordinates $(\eta, \xi, \tau)$ for $\epsilon <<1$, by defining, 
\begin{equation}
	\xi=\epsilon^{1/2} (y - Ut); \eta = \epsilon^{1/2} x
\end{equation}
and 
\begin{equation}
	\tau = \epsilon^{3/2} t
\end{equation}
which give $\partial_y =\epsilon^{1/2} \partial_{\xi}$, $\partial_x = \epsilon^{1/2} \partial_{\eta}$ and $\partial_t = (- \epsilon^{1/2} U \partial_{\xi} + \epsilon^{3/2} \partial_{\tau})$ where the wave phase speed $U_0$ is normalized with $v_{De}$ and hence $U=\frac{U_0}{v_{De}}=1$. Note that the ion velocity in normalized form can be expressed as,
\begin{equation}
	{\bf v}_{i \perp}= - \frac{L_n}{\rho_s} \nabla_{\perp} \Phi \times \hat{z} - (\partial_t + {\bf v}_E \cdot \nabla_{\perp}) \nabla_{\perp} \Phi ={\bf v}_E + {\bf v}_P
\end{equation}
The expansion of physical quantities in terms of $\epsilon$ are defined as follows,
\begin{equation}
	n_i = 1+ \epsilon n_{i1} +\epsilon^2 n_{i2} + ...
\end{equation}
\begin{equation}
	\Phi = \epsilon \Phi_1 + \epsilon^2 \Phi_2 + ...
\end{equation}

\begin{equation}
	\frac{\tilde{n_i}}{n_{i0}} \simeq \epsilon \Phi_1 + \epsilon^2 (\frac{1}{2} \Phi_1^2 + \Phi_2)
\end{equation}
where quasi-neutrality $n_i=n_e$ has been used.
Now we analyze each term of Eq. (12) in different order in stretched coordinates and write the expressions only up to order $\epsilon^{5/2}$ as follows.
\begin{equation}
	\begin{aligned}
		&\partial_t \frac{\tilde{n_i}}{n_{i0}} \simeq (- \epsilon^{1/2} U \partial_{\xi} + \epsilon^{3/2} \partial_{\tau}) [ \epsilon \Phi_1 + \epsilon^2 (\frac{1}{2} \Phi_1^2 + \Phi_2) ]\\&
		=\epsilon^{3/2}[- U \partial_{\xi} \Phi_1] + \epsilon^{5/2} [\partial_{\tau} \Phi_1 - U \partial_{\xi} (\frac{1}{2} \Phi_1^2 + \Phi_2)]
	\end{aligned}
\end{equation}

\begin{equation}
	\begin{aligned}
		&\partial_t \nabla_{\perp}^2 \Phi \simeq - (- \epsilon^{1/2} U \partial_{\xi} + \epsilon^{3/2} \partial_{\tau}) ( \epsilon \partial_{\eta}^2 + \epsilon \partial_{\xi}^2) (\epsilon \Phi_1 + \epsilon^2 \Phi_2) \\&
		\simeq \epsilon^{5/2} [U( \partial_{\eta}^2 +  \partial_{\xi}^2) \partial_{\xi} \Phi_1]
	\end{aligned}
\end{equation}

\begin{equation}
	\begin{aligned}
		&(\nabla_{\perp} \Phi \times \hat{z}. \hat{x}) \partial_y \Phi = \epsilon^{1/2} \partial_{\xi} [\epsilon \Phi_1 + \epsilon^2 \Phi_2]= \epsilon^{3/2}[\partial_{\xi} \Phi_1] \\&
		+ \epsilon^{5/2} [\partial_{\xi} \Phi_2]
	\end{aligned}
\end{equation}
Let us look at the following term,
\begin{equation}
	\begin{aligned}
		&\frac{L_n}{\rho_s} (\nabla_{\perp} \Phi \times \hat{z} \cdot \nabla_{\perp}) \nabla_{\perp}^2 \Phi= \frac{L_n}{\rho_s} (\partial_y \Phi \partial_x - \partial_x \Phi \partial_y) \nabla_{\perp}^2 \Phi \\&
		=\frac{L_n}{\rho_s} \{ \Phi, \nabla_{\perp}^2 \Phi \} 
	\end{aligned}
\end{equation}
where $\{ \Phi, \nabla_{\perp}^2 \Phi \} = (\partial_y \Phi \partial_x \nabla_{\perp}^2 \Phi- \partial_x \Phi \partial_y \nabla_{\perp}^2 \Phi)$ is Poisson bracket and,
\begin{equation}
	\begin{aligned}
		&\{ \Phi, \nabla_{\perp}^2 \Phi \} = \epsilon^{1/2} \partial_{\xi} (\epsilon \Phi_1 + \epsilon^2 \Phi_2) \epsilon^{1/2} \partial_{\eta} ( \epsilon \partial_{\eta}^2 + \epsilon \partial_{\xi}^2) \\&
		(\epsilon \Phi_1 + \epsilon^2 \Phi_2) - \epsilon^{1/2} \partial_{\eta} (\epsilon \Phi_1 + \epsilon^2 \Phi^2) \\&
		\epsilon^{1/2} \partial_{\xi} (\epsilon \partial_{\eta}^2 + \epsilon \partial_{\xi}^2) (\epsilon \Phi_1 + \epsilon^2 \Phi_2)
	\end{aligned}
\end{equation}
In stretched coordinates, we define $\nabla_{\perp}=(\partial_{\eta}, \partial_{\xi})$ and  $\nabla_{\perp}^2=(\partial_{\eta}^2 + \partial_{\xi}^2)$ and hence the above relation in the lowest order becomes,
\begin{equation}
	\{ \Phi, \nabla_{\perp}^2 \Phi \}= \epsilon^4 \{ \Phi_1, \nabla_{\perp}^2 \Phi_1 \}
\end{equation}
Equations (30) and (28) imply,
\begin{equation}
	\frac{L_n}{\rho_s} (\nabla_{\perp} \Phi \times \hat{z} \cdot \nabla_{\perp}) \nabla_{\perp}^2 \Phi = \epsilon^4 \frac{L_n}{\rho_s} \{ \Phi_1, \nabla_{\perp}^2 \Phi_1 \}
\end{equation}
Using the relations (25, 26, 27) and ((31), the Eq. (12) can be expressed in the following form,
\begin{equation}
	\begin{aligned}
		&\epsilon^{3/2} [-U \partial_{\xi} \Phi_1 + \partial_{\xi} \Phi_1 ] + \epsilon^{5/2} [\{ \partial_{\tau} \Phi_1  - U \partial_{\xi} (\frac{1}{2}\Phi_1^2 + \Phi_2) \} \\&
		+ \partial_{\xi} \Phi_2 + U \partial_{\xi} \nabla_{\perp}^2 \Phi_1] 
		+ \epsilon^4 \frac{L_n}{\rho_s} \{ \Phi_1, \nabla_{\perp}^2 \Phi_1 \} =0
	\end{aligned}
\end{equation} 
If we assume $\frac{L_n}{\rho_s}$ of the order of $\epsilon^{-3/2}$, then in order $\epsilon^{5/2}$, we obtain,
\begin{equation}
	\begin{aligned}
		& [\{ \partial_{\tau} \Phi_1  - U \partial_{\xi} (\frac{1}{2}\Phi_1^2 + \Phi_2) \} + \partial_{\xi} \Phi_2 + U \partial_{\xi} \nabla_{\perp}^2 \Phi_1] \\& 
		+ \frac{L_n}{\rho_s} \{ \Phi_1, \nabla_{\perp}^2 \Phi_1 \}  \simeq 0
	\end{aligned}
\end{equation}
where $[-U \partial_{\xi} \Phi_1 - \partial_{\xi} \Phi_1 ]=0$ for $U=1$ and the above equation reduces to,
\begin{equation}
	[ \partial_{\tau} \Phi  -  \partial_{\xi} (\frac{1}{2}\Phi^2)  +  \partial_{\xi} \nabla_{\perp}^2 \Phi] 
	+ \frac{L_n}{\rho_s} \{ \Phi, \nabla_{\perp}^2 \Phi \} \simeq 0
\end{equation}
while $\Phi_1= \Phi$ has been used.
The vorticity term can be of the same order as the scalar nonlinear term  ($\frac{1}{2} \Phi^2$) if $\frac{L_n}{\rho_s} \simeq \epsilon^{-3/2}$ that is $\frac{\rho_s}{L_n} \simeq \epsilon^{3/2}$. Equation (34) is the general nonlinear equation for pure drift waves in stretched coordinates $(\eta, \xi, \tau)$. If $\epsilon^4 \frac{L_n}{\rho_s}<<\epsilon^{5/2}$, then the Eq. (34) in order $\epsilon^{5/2}$ becomes,
\begin{equation}
	[ \partial_{\tau} \Phi  -  \partial_{\xi} (\frac{1}{2}\Phi^2)  +  \partial_{\xi} \nabla_{\perp}^2 \Phi] 
	 \simeq 0
\end{equation}

The solution of nonlinear Eq. (34) is not straightforward because it contains both the effects of compressibility via electron scalar nonlinear term \cite{Saleem2024} and ion vorticity term which was considered by Hasegawa and Mima \cite{hasegawa1978pseudo}. Another new equation (35) can be obtained by simply ignoring the higher order vorticity term in Eq. (34). The Hasegawa-Mima equation can be obtained if the nonlinear electron density term is dropped. Both these limiting cases are discussed in the following sections for the sake of completeness. 

\section{Soliton solution}
The solitary structures formed by the nonlinear drift waves were first investigated by Nozaki and Tanuiti \cite{nozaki1974propagation} and later on by Gell \cite{gell1977drift}. These authors derived the modified KdV equations for nonlinear drift waves. But they normalized the ion velocity with ion acoustic speed $c_s$. However, they also used the Boltzmann distribution for electrons approximating $n_i\simeq n_e \simeq (\Phi + \frac{1}{2} \Phi^2)$. It has been pointed out that the phase speed of the drift wave is electron diamagnetic drift $v_{De}$, therefore the ion velocity should be normalized with $v_{De}$ in the derivation of KdV equation for drift waves . For details, the reference \cite{Saleem2024} may be consulted.
Equation (35) can be solved to obtain a two-dimensional bounded structure by defining a coordinate $\chi = \chi (\eta, \xi, \tau)$ moving with constant speed say, M, as,
\begin{equation}
\chi= \lambda \eta + \xi - M \tau
\end{equation}
where $\lambda <1$ because the wave propagates predominantly along $\xi$ direction. Then Eq. (35) in $\chi$ frame becomes,
\begin{equation}
M d_x \Phi - \frac{1}{2} d_x \Phi^2 + (1+ \lambda^2) d_x^3 \Phi =0
\end{equation}
where $d_x=\frac{d}{d x}$. Using boundary conditions $(\Phi, d_x \Phi, d_x^2 \Phi) \rightarrow 0$ as $\chi \rightarrow \pm \infty$, the above equation can be expressed as,
\begin{equation}
(\Phi^{\prime})^2 = \frac{M \Phi^2 + \frac{1}{3} \Phi^3}{(1+ \lambda^2)}
\end{equation}
where $\Phi^{\prime}= d_x \Phi$. This equation admits the solution in following form with negative $\Phi$,
\begin{equation}
\Phi (\chi)= -3M \sech^2 (a \chi)
\end{equation}
where $a =\frac{\sqrt{M}}{2 \sqrt{(1+ \lambda^2)}}$. For illustration, the solution has been plotted in Fig. (1) for $M=0.1$ and $\lambda=0.1$.

If $\Phi=\Phi(y,t)$ is assumed, then the vorticity term disappears in Eq. (34) automatically and it reduces to Eq. (5) because $\nabla_{\perp}^2=\partial_{\xi}^2$ in this case. In moving coordinate $\chi= \xi - M \tau$
where $\frac{M_0}{v_{De}}$ is normalized speed of the solitary structure,  
the solution of Eq. (5) turns out to be,
\begin{equation}
	\Phi(\chi) = - \mid \Phi_m \mid \sech^2 (\frac{\chi}{W})
\end{equation}
where the maximum normalized amplitude is $\mid \Phi_m \mid=3M<1$ and width of the solitary structure is $W= \sqrt{\frac{4}{M}}$. Equations (5) and (40) are the same as equations (34) and (36) of Ref. \cite{Saleem2024}. The above KdV equation creates dip soliton while the hump soliton is produced by the KdV equation for ion acoustic waves. 
\begin{figure}\label{scalghz1}
    \centering
    \includegraphics[width=8cm, height=8cm]{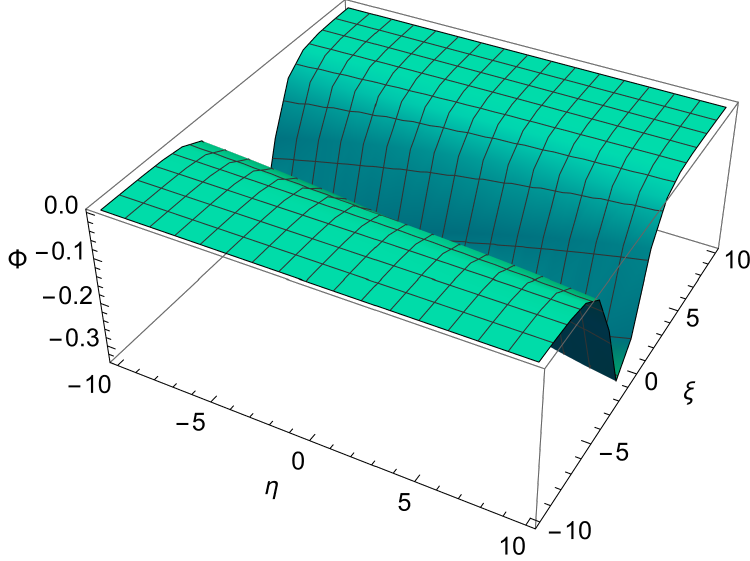}
    \caption{The 3D plot of negative electrostatic potential $\Phi$ vs moving 
coordinate $\chi$ is shown for $M=0.1$ and $\lambda = 0.1$.}
\end{figure}

\section{Vortex solution}
In our point of view, Eq. (13) does not yield the HM Eq. (10) because the scalar nonlinear term is not negligible, in general. For the sake of completeness, we explore what happens if this term is ignored.
In this case, we get from Eq. (34), an equation similar to HM equation in stretched coordinates as follows,
\begin{equation}
	\frac{\rho_s}{L_n}[\{ \partial_{\tau} \Phi   +  \partial_{\xi} \nabla_{\perp}^2 \Phi] 
	- \{\nabla_{\perp}^2 \Phi, \Phi \} =0
\end{equation}
In form, this is somewhat different from standard HM equation (10) because it has been obtained in stretched coordinates assuming $U=1$. However, it admits a solution in the form of dipolar vortex similar to HM equation. For simplicity we denote the planar coordinates $(\eta, \xi)$ by $(X,Y)$ in this section. Let us define a moving coordinate $Y=\xi - V \tau$ where $V=\frac{V_0}{v_{De}}$ and $V_0$ is the constant speed of the vortex which gives $\partial_{\xi}=\partial_Y$ and $\partial_{\tau}= - V \partial_Y$. Let $a_0=- \frac{\rho_s}{L_n}$ and $b_0= V \frac{\rho_s}{L_n}$. Then (41) can be expressed as,
\begin{equation}
	a_0 \partial_Y \nabla_{\perp}^2 \Phi + b_0 \partial_Y \Phi + \{\nabla_{\perp}^2 \Phi, \Phi \} =0
\end{equation}
The equation (42) can also be written as,
\begin{equation}
	\{(\nabla_{\perp}^2 \Phi- \lambda^2 \Phi), (\Phi +a_0 X) \} =0
\end{equation}
where $\lambda^2=(-\frac{b_0}{a_0})>0$.
The above equation is satisfied if $(\nabla_{\perp}^2 \Phi-\lambda^2 \Phi)$ is a function of 
$(\Phi +a_0 X)$.  Let us choose the simplest case,
\begin{equation}
	\nabla_{\perp}^2 \Phi-\lambda^2 \Phi= C_0 (\Phi +a_0 X)
\end{equation}
where $C_0$ is a constant to be determined. The above equation can be expressed as,
\begin{equation}
	\nabla_{\perp}^2 \Phi-\mu^2 \Phi= d_0 X
\end{equation}
where $\mu^2=(C_0 + \lambda^2)$ and $d_0=C_0 a_0$.
Now we transform the above equation from $(X,Y)$ to polar coordinates $(r, \theta)$ with $X=r \cos \theta$, $Y=\sin \theta$, and $\theta = \tan^{-1} \frac{Y}{X}$. In polar coordinates, Eq. (45) becomes,
\begin{equation}
	\nabla_{\perp}^2 \Phi- \mu^2 \Phi= d_0 (r \cos \theta)
\end{equation}
We consider a circle of radius $R_0$ and find the solutions of Eq. (46) in inner $r<R_0$ and outer $r>R_0$ regions of this circle following Ref. \cite{larichev1976two}. Let us use the variable separation technique by defining $\Phi(r, \theta)= \psi (r) f(\theta)$ and $f(\theta) = \cos \theta$. Equation (46) takes the form,
\begin{equation}
	r^2 \partial_r^2 \Phi + r \partial_r \Phi -(\mu^2 r^2 +1) \Phi =d_0 r^3 \cos \theta
\end{equation}
Let $s_0^2=- \mu^2$ and $r^{\prime} =s_0 r$, then Eq. (46) becomes,
\begin{equation}
	r^{\prime 2} \frac{d^2}{d r^{\prime 2}} \psi (r^{\prime}) +r^{\prime} \frac{d}{d r^{\prime}}\psi (r^{\prime}) +(r^{\prime 2} -1)\psi (r^{\prime})= d_0 \frac{r^{\prime 3}}{s_0^3}
\end{equation}
Solution of above equation in the inner region turns out to be,
\begin{equation}
	\Phi_{In} (r, \theta) = (P_1 J_1 (s_0 r) + \frac{r d_0}{s_0^2}) \cos \theta
\end{equation}
where $P_1$ is a constant and $J_1 (s_0 r)$ is Bessel function of order one.
The outer solution of (47) turns out to be,
\begin{equation}
	\Phi_{out} = Q_1 K_1(\lambda r) \cos \theta
\end{equation}
where $Q_1$ is a constant and $K_1(\lambda r)$ is modified Bessel function of order one. The continuity condition for $\Phi$ requires that the inner and outer solutions and their derivatives must be equal at $r=R_0$. Using continuity conditions for $\Phi$ at $r=R_0$, one obtains,
\begin{equation}
	P_1 = - \frac{\lambda^2}{(\lambda^2 + s_0^2)} [\frac{R_0 d_0}{s_0^2 J_1 (s_0 R_0)}]
\end{equation}
and,
\begin{equation}
	Q_1 =  \frac{s_0^2}{(\lambda^2 + s_0^2)} [\frac{R_0 d_0}{s_0^2 K_1 (\lambda R_0)}]
\end{equation}
The only unknown is now the parameter $s_0$ which can be estimated numerically using the relation,
\begin{equation}
	\frac{K_1 (\lambda R_0)}{K_2 (\lambda R_0)}= - \frac{s_0}{\lambda} \frac{J_1 (s_0 R_0)}{J_2 (s_0 R_0)}
\end{equation}
where $J_2 (s_0 R_0)$ is the Bessel function of order two and $K_2 (\lambda R_0)$ is modified Bessel function of order two. The solution of HM like equation has been discussed in several research articles and books.

\section{Summary}
A geeneral nonlinear equation for two-dimensional drift waves has been derived to demonstrate that the Hasegawa-Mima (HM) equation does not fully capture the dynamics of these waves. Specifically, the electron nonlinear term $\frac{1}{2} \Phi^2$ must also appear along with the ion vorticity term in the nonlinear equation for drift waves. Since Hasegawa and Mima~\cite{hasegawa1978pseudo} normalized the density term in ion continuity equation with the total density $n_i= n_0(x) + \tilde{n_i}(x,y,t)$ and used $\frac{\partial_t n_i}{n_i}=\partial_t \ln n_i= \partial_t \Phi$, therefore the nonlinear electron density term was artificially killed. 

In our opinion, the correct expression is $\frac{{n}_i}{n_{i0}} = e^{\Phi}$ which yields $\frac{\tilde{n}_i}{n_{i0}} \simeq (\Phi+ \frac{1}{2} \Phi^2)$. It may be mentioned here that Washim-Tanuiti \cite{washimi1966propagation}, Saleem \cite{Saleem2024} and many other authors follow the same relation to derive nonlinear equations for low frequency electrostatic waves. This correction introduces a scalar nonlinear term, $\frac{1}{2} \Phi^2$, in the Hasegawa-Mima equation. Then the complete nonlinear equation for drift waves takes the form of Eq. (13).

Interestingly, in the drift wave’s phase velocity frame, the scalar nonlinearity from electrons Boltzmann distribution and the vector nonlinearity from ion vorticity are inherently coupled. If the electrostatic potential depends only on a single coordinate—as in the chosen geometry $\Phi = \Phi(y)$—then the electron nonlinearity leads to the formation of density cavities~\cite{Saleem2024}. In contrast, if the electron nonlinear contribution is suppressed and a two-dimensional potential $\Phi = \Phi(x, y)$ is assumed, the ion vorticity generates dipolar vortices, similar to the solution of Hasegawa-Mima equation. Furthermore, when the ion vorticity is assumed to be a small effect, Eq. (13) gives a two dimensional equation (35) which admits solution in the form of two dimensional soliton shown in Fig. (1).
From the author's perspective, the electron nonlinearity is the dominant effect and should not be neglected in the derivation of nonlinear drift wave equation.

The key conclusions of this work can be summarized as follows:
\begin{itemize}
    \item[\textbf{I.}] The general nonlinear equation (13) for pure drift waves has been derived incorporating both nonlinear terms $\frac{1}{2} \partial_t \Phi^2$ and $\frac{L_n}{\rho_s} (\nabla_{\perp} \Phi \times \hat{z} \cdot \nabla_{\perp}) \nabla_{\perp}^2 \Phi$.  Equation (13) reduces to Hasegawa-Mima equation (10) if the term $\partial_t ( \frac{1}{2} \Phi^2)$ is ignored.
    
     \item[\textbf{II.}] The nonlinear electron density term is generally dominant over the ion vorticity term particularly for $\rho_s^2 k_{\perp}^2<1$. 
     
    \item[\textbf{III.}] The nonlinear term $\frac{1}{2} \partial_t \Phi^2$ is missing in the original Hasegawa-Mima equation because they approximated $\frac{\partial_t \tilde {n_i}}{n_{i0}}=\partial_t \Phi$ whereas the quasi-neutrality condition yields $\frac{\partial_t \tilde {n_i}}{n_{i0}} \simeq \partial_t (\Phi + \frac{1}{2} \Phi^2)$ assuming the inertia-less electrons to follow the Bltzmann distribution. The nonlinear equation (13) can yield solitons in one limit (when ion vorticity is ignored) and vortices in another limit (when nonlinear density term is ignored).
\end{itemize}

This work is expected to be very useful for future studies of nonlinear drift waves which have wide-ranging applications in space, astrophysical, and laboratory plasmas.

\section{Data Availability Statement}
The data used for the preparation of the presented results is included in the text of this work.

\bibliographystyle{unsrt} 
\bibliography{bibliography} 

\end{document}